\documentclass{sig-alternate}

\usepackage{algorithm}
\usepackage{algorithmic}

\long\def\comment#1{}
\begin{document}

\title{Hybrid Spam Filtering for Mobile Communication}

\numberofauthors{3} 
\author{
\alignauthor
Ji Won Yoon\\
       \affaddr{Robotics Research Group}\\
       \affaddr{Engineering Department,}\\
       \affaddr{University of Oxford, UK}\\
       \email{jwyoon@robots.ox.ac.uk}
\alignauthor 
Hyoungshick Kim\\
       \affaddr{Security Group}\\
       \affaddr{Computing Laboratory,}\\
       \affaddr{University of Cambridge, UK}\\
       \email{hk331@cl.cam.ac.uk}
\alignauthor Jun Ho Huh\\
       \affaddr{Software Engineering Group}\\
       \affaddr{Computing Laboratory,}\\
       \affaddr{University of Oxford, UK}\\
       \email{jun.ho.huh@comlab.ox.ac.uk}
}

\maketitle

\begin{abstract}
Spam messages are an increasing threat to mobile communication. Several
mitigation techniques have been proposed, including white and black listing,
challenge-response and content-based filtering. However, none are perfect
and it makes sense to use a combination rather than just one. We propose an anti-spam framework based on 
the hybrid of content-based filtering and challenge-response. There is the trade-offs between 
\emph{accuracy} of anti-spam classifiers and the \emph{communication overhead}. 
Experimental results show how, depending on the proportion of spam messages, different filtering 
parameters should be set.
\end{abstract}

\comment{

}

\section{Introduction}
\label{section: introduction}

Short Message Service (SMS) and Multimedia Messaging Service (MMS) are a popular means of mobile 
communication. Texting costs have decreased continuously over the years (to an extent of free texting) 
whereas the bandwidth for communication has increased dramatically. Such trends have attracted a large 
number phishing and spamming attacks using text messages. In particular, spam messages containing 
pornographic or promotive materials are an emerging phenomenon, and they have caused a significant level 
of inconvenience for users. These are now prevalent in Korea, Japan and China and prone to spread across 
countries where mobile communication is popular. Statistics for 2008 show that a user in China, on 
average, receives 8.29 SMS spam per week \cite{He08:Challenge}. 

Much of the existing research into anti-spam solutions, however, has focused on the protection of emails 
in the context of the Internet. Some of the popular methods include white and black listing, digital 
signature, postage control, address management, collaborative and content-based filtering 
\cite{Healy05:Spam, Cormack207:Spam, Dwork03:Spam, Hall98:Spam, Golbeck04:Spam}. Different 
characteristics between emails and text messages make it harder for one to apply such approaches directly 
in mobile networks and analyze the results. Each approach has its own set of drawbacks and does not 
improve much if used alone. For example, the extra traffic required to perform challenge-response needs 
to be minimized (or needs to be compensated for) as it is more expensive to use the bandwidth in mobile 
networks. This issue is capable being addressed by content-based filtering: obvious spam would be 
filtered first to reduce the number of messages subject to challenge-response. In this paper, we propose 
an anti-spam framework based on the combination of these two methods. We attempt to reduce a great number 
of high-volume spamming, and as a result, minimize the extra amount of bandwidth that would be required. 
Given a reasonable filtering algorithm, we show that, ultimately, less bandwidth (than freely allowing 
high-volume spam) will be used with our method. 

The remainder of the paper is organized as follows. In Section \ref{section: Proposed Systems}, we 
describe a hybrid spam filtering framework. Section \ref{section: Evaluation} evaluates the performance 
of our hybrid method based on two measures, the traffic usage and the accuracy. Finally, in Sections 
\ref{section: conclusion} we discuss the contribution of this paper and the remaining work.  

\section{Existing Solutions}

Content-based filtering solutions have been proved to be effective against emails, which are typically 
larger in size compared to text messages. Abbreviations and acronyms are used more frequently in SMS and 
they increase the level of ambiguity. \cite{Hidalgo06:Spam} propose binary classification and filtering 
methods for short messages in a Bayesian scheme. Classification rules are defined and extended from 
general patterns identified in past spam. However, adaptive schemes as such are weak against innovative 
attacks where strategies constantly evolve to manipulate classification rules. Filtering alone will not 
be sufficient to detect spam.

Many anti-spam solutions \cite{He08:Challenge, Shirali07:Challenge} have been suggested based on a 
challenge-response protocol. A message sender needs to verify that they are a legitimate sender by 
answering the challenge message (e.g. through a web interface) before their message is forwarded to the 
recipient. The sender authenticates themselves as a human-user by answering a simple turing test for 
which a machine cannot easily understand. Nevertheless, the protocol has often been criticized for extra 
user interaction and traffic used. There might also be a significant overhead in storing and managing 
challenge messages.

\comment{
methods. Firstly, mobile users tend to believe that their devices are always connected and the sent 
messages will be delivered to the correct recipient on time. Anti-spam solutions, however, are imperfect 
and may classify a legitimate message as spam; it may be rejected and moved to a spam folder or returned 
to the sender. 

majority of users would not want to miss the legitimate messages even if this means manual filtering. It 
is more expensive to send a text message than an email. Operators usually charge per-packet regardless of 
the content. Hence, the traffic usage needs to be carefully considered upon designing an anti-spam 
solution for mobile networks.
}

Our goal is to develop a solution that ultimately minimizes the usage of network bandwidth by 
discouraging high-volume spamming. We believe the extra traffic required to perform challenge-response 
can be compensated if a large amount of spamming attacks can be reduced as a result.
In our approach the challenge-response protocol classifies machine-generated spam. We also use the 
filtering method to reduce the number of messages that need to be verified. Simulation results in later 
sections show that this hybrid approach is capable of controlling high-volume spam and the traffic usage.

\section{A Hybrid Framework}
\label{section: Proposed Systems}

Text messages are classified into three different regions using a content-based filtering method: normal, 
uncertain and spam. A filtering method cannot deal with \emph{uncertain} messages; therefore, we use a 
challenge-response protocol to classify the uncertain messages into normal and spam regions (see Fig. 1). 
We assume that the majority of spam messages are generated by machines.

A human verification mechanism (in the form of challenge-response) is added to a common filtering scheme 
to detect whether an uncertain message falls into the normal or the spam region. A message center (owned 
by an operator) sends a challenge query to check if the sender is an individual or a machine. The sender 
responds by answering the query and the center compares the returned value against the known correct 
value. If the values match, the message is classified as \emph{normal}, else, as \emph{spam}. We are 
interested in further classification of this uncertain region.
\begin{figure}[!h]
\centering
\includegraphics[scale=0.2]{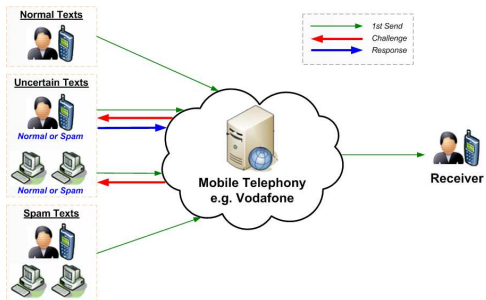}
\caption{Hybrid Spam Filtering Overview}
\label{fig: frameworks in mobile system}
\end{figure}
A message center is given the full responsibility of running the framework due to the following reasons. 
Firstly, it should reduce the traffic usage by filtering spam as early as possible, before forwarding 
them to the recipient. Secondly, using the challenge-response protocol, the center will be able to 
collect an enormous amount of sample data in real time; these can be used to develop highly effective 
classifiers and continuously improve the performance of filtering algorithms. Lastly, it would be 
difficult to install and maintain a homogeneous anti-spam software on all mobile devices; instead we rely 
on one solution deployed in a message center.
\subsection{Uncertain Region}
\label{section: Three labels with uncertain area}
\begin{figure}[!h]
\centering
\begin{tabular}{c c}
\includegraphics[scale=0.27]{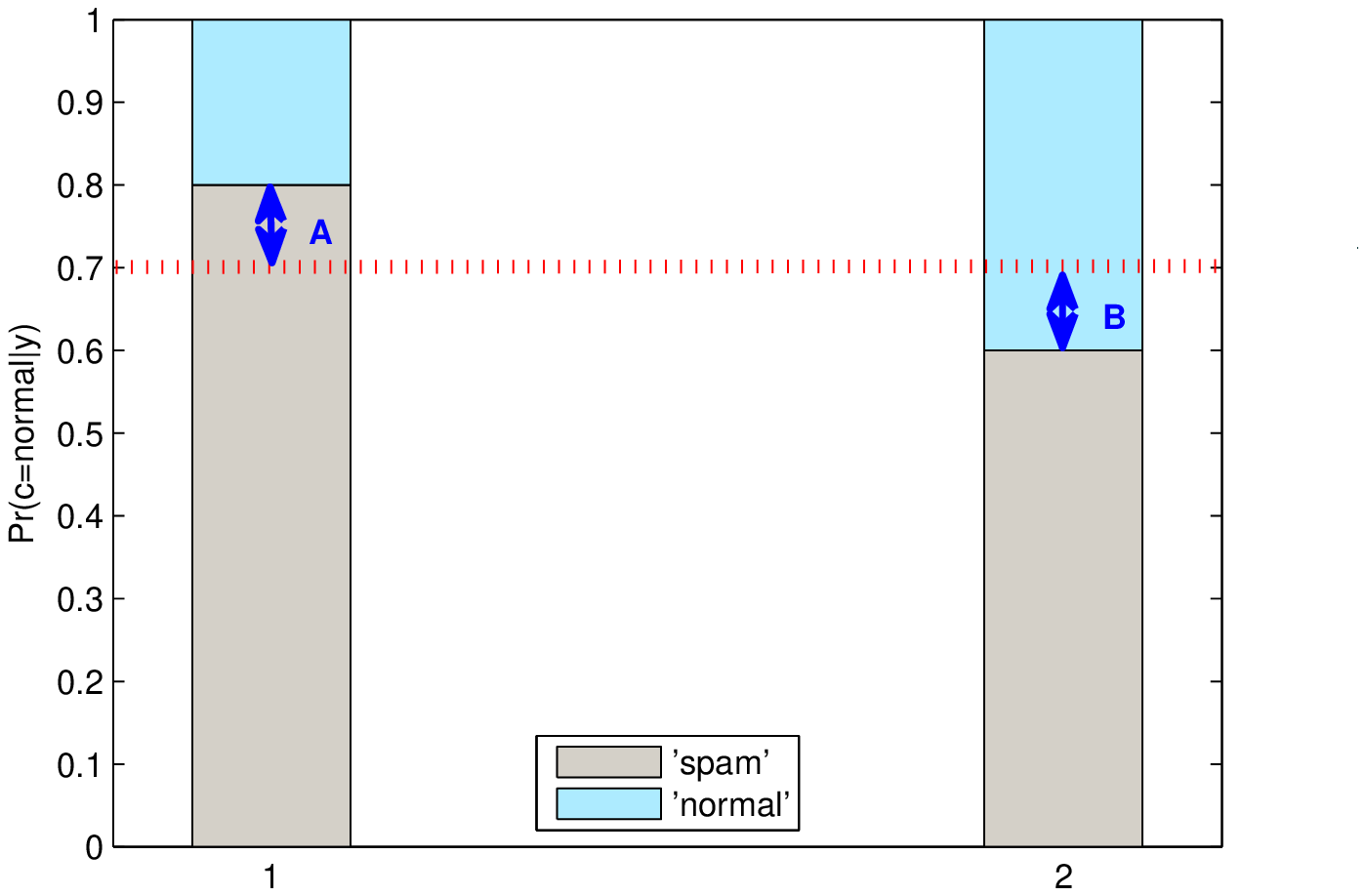}&
\includegraphics[scale=0.27]{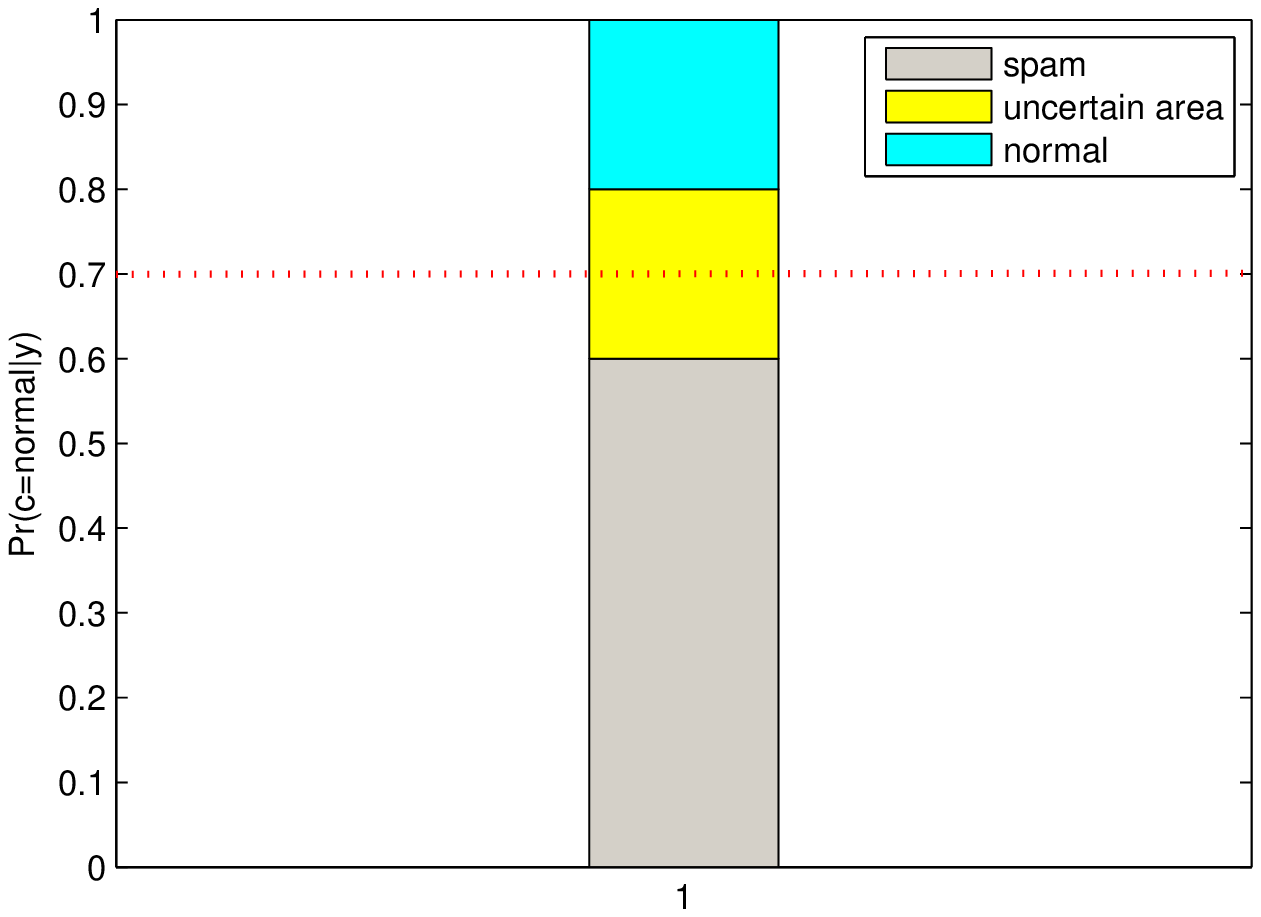}\cr
(a) Two possible cases&
(b) Uncertain regions
\end{tabular}
\caption{(a) Two possible cases: $h>\tilde{h}$ (case 1) and $h<\tilde{h}$ (case 2) for a given ground 
truth $\tilde{h}$ (red dot line) and (b) Modified classification embedding uncertain area given a ground 
truth $\tilde{h}$ (red dot line)}
\label{fig: Threshold based approach}
\end{figure}
If we assume there are only two regions, normal and spam, a filtering method will use binary 
classification. Suppose that we have a probabilistic model for the anti-spam classifier as a posterior 
distribution $Pr(c=\textrm{normal}|y)$. This is the probability that a message is normal: $c$ and $y$ 
denote realization of random variables for a class and a message, respectively. The odd ratio of the 
posterior is used to obtain a measurable classification by $O_{post} = 
\frac{Pr(c=\textrm{normal}|y)}{Pr(c=\textrm{spam}|y)}$. If $O_{post}>1$, a message is classified as 
normal; otherwise, as spam. Alternatively, we can simply use a threshold based approach in the posterior 
distribution. If $Pr(c=\textrm{\emph{normal}}|y)$ is close to one, a message is likely to be normal; if 
close to zero, it is likely to be spam. Let $\bar{c}=f(y, h)$ be a spam filter where $\bar{c}$ and $h$ 
are an output and a given threshold, respectively. This filter would work with the following rules:
\begin{equation}
\bar{c} = f(y, h) = \left\{
\begin{array}{c c}
\textrm{normal}	& \textrm{if } Pr(c=\textrm{normal}|y) \geq h\cr
\textrm{spam}		& \textrm{if } Pr(c=\textrm{normal}|y) < h
\end{array}\right.
\end{equation}
This separates normal messages from spam (odd ratio approach is a special case where $h=0.5$). The main 
problem with this approach is finding a proper threshold; because the threshold for ground truth 
$\tilde{h}$ is unknown, there are two possible cases as shown in Fig. \ref{fig: Threshold based 
approach}. If $h$ is higher than $\tilde{h}$, some of the normal messages in region \emph{A} may be 
classified as spam. If $h$ is lower than $\tilde{h}$, some of the spam in region \emph{B} may bypass the 
filter and reach the recipients. Such a threshold problem will always be present in classification: it is 
almost impossible to find the underlying $\tilde{h}$; and anti-spam software companies are likely to use 
strategies based on their own experiences. The sensitivity problem can be resolved by introducing an 
uncertain region with two thresholds (see Fig. \ref{fig: Threshold based approach}-b). These can be 
implemented as the upper and lower boundaries of a traditional threshold system. There are three labels: 
spam, uncertain area, and normal; we focus on the uncertain area. Spam and normal regions are classified 
as in the traditional system. Only the messages that fall into the uncertain area are checked using 
challenge-response. In the next section we describe the protocols in detail.  

\subsection{Challenge-Response Protocols}
\label{section: Challenge-response protocol}

We assume that there is a turing test available with a low probability of producing false positives and 
negatives. Completely Automatic Public Test to tell Computer and Humans Apart (CAPTCHA) is a commonly 
used one: it generates pattern matching problems for which a human can easily recognize and a machine 
cannot. An automated program that generates thousands of spam messages will not be capable of answering a 
CAPTCHA based challenge, which might be a graphical image containing a faint typeface. If the response is 
correct, there is a high probability that the sender is a human. A CAPTCHA can be designed in a flexible 
manner using different media forms such as an image, an audio file and a text \cite{Ahn08:CAPTCHA}. Their 
implementation details are beyond the scope of this paper. 

A number of challenge-response protocols have already been proposed \cite{He08:Challenge, 
Shirali07:Challenge}. However, these focus only on the implementation issues without considering the 
security model and the cryptographic details. This section defines our security models and describes a 
number of possible protocols in line with them. There are several issues we need to consider before 
designing the protocols. Firstly, in dealing with spam, message authentication and integrity are 
important; whereas, confidentiality is not. Secondly, text messages are usually unencrypted and unsigned; 
it is possible to tamper with them during transmission. Thirdly, security properties of the communication 
channel between a message center and a sender need to be defined; this channel might or might not be an 
authenticated one. Lastly, management of the session information between all trusted pairs while 
performing challenge-response, would impose a huge storage overhead on a message center; there might be 
more than one message center sharing this information; and it might or might not be stored in the center. 
Mindful of these security and scalability issues, we propose four different protocols: protocols 3 and 4 
have been designed with the assumption of an authenticated channel, and protocols 1 and 2 have not; 
moreover, protocols 1 and 3 assume that a message center manages the session information, and the others 
do not.  

\subsubsection{Notations}
The symbols $S$ and $R$ represent a sender and a recipient, respectively. $M$ represents a mobile message 
center, $T$ a timestamp, $N$ a nonce, $K$ a key and $K^{-1}$ its inverse. In a symmetric crypto-system 
such as AES, $K$ and $K^{-1}$ are always equal. We use $\{P\}_K$ for a plain text message $P$ encrypted 
with $K$. $H$ is a one-way hash function. The subscript $m$ in $K_m$ implies that $K_m$ is $M$'s public 
key. In addition, $ms$ in $K_{ms}$ shows that $K_{ms}$ is intended for communication between $M$ and $S$.

A sender's ability to respond to a challenge depends on knowing and interpreting a key, ${K_c}^{-1}$. A 
non-authorized sender (e.g. a program sending spam) will not be able to interpret and gain information 
about ${K_c}^{-1}$; this key serves to identify machine-generated spam. For simplicity, encryption 
algorithms are not considered in our protocols. 

\subsubsection{Protocols}
In protocol 1, the message center, $M$, maintains the session information.

\textbf{[Protocol 1]}\\
\begin{tabular}{ll}
(M1) $S \longrightarrow M: $&$ S, R, P$\\
(M2) $M \longrightarrow S: $&$ M, S, \{K_{ms}\}_{K_c}, \{H(S, R, P), N\}_{K_{ms}}$\\
(M3) $S \longrightarrow M: $&$ S, M, \{H(S, R, P), N + 1\}_{K_{ms}}$\\
\end{tabular}

Before sending message 1, $S$ stores $R$ and $P$ to prevent message modification attacks. After receiving 
message 1, $M$ generates $K_{ms}$ and stores $(S, R, P, K_{ms}, N)$ as the session information. $K_{ms}$ 
is protected with ${K_c}$. An image CAPTCHA would be one way of protecting $K_{ms}$ against spam 
programs. After receiving message 2, $S$ decrypts $\{K_{ms}\}_{K_c}$ by answering the challenge (their 
ability to interpret ${K_c}^{-1}$). $S$ then decrypts $H(S, R, P)$ and $N$ using $K_{ms}$. $S$ compares 
$H(S, R, P)$ against the previously stored values. $S$ terminates the protocol if these values do not 
match; otherwise, $S$ generates $\{H(S, R, P), N + 1\}_{K_{ms}}$ by $K_{ms}$ and sends it to $M$. After 
receiving message 3, $M$ verifies $\{H(S, R, P), N + 1\}_{K_{ms}}$. If it is valid, $M$ forwards the 
stored message $(S, R, P)$ to $R$. Finally, $M$ deletes the session information.

Users will be frustrated if challenge-response happens too often. We use a timestamp, $T$, to solve this 
problem. After receiving message 3, $M$ maintains a session information $(S, R, P, K_{ms}, T)$ between 
$S$ and $R$ for a given time interval. $M$ checks the validity of $K_{ms}$ using the session information 
and a policy that defines the lifetime of $K_{ms}$.

The main drawback of this protocol is that $M$ has to bear the huge overhead of maintaining the session 
information. We describe another protocol which solves this issue by using authorized tokens 
instead:\newline\newline
\textbf{[Protocol 2]}\\
\begin{tabular}{ll}
(M1) $S \longrightarrow M: $&$ S, R, P$\\
(M2) $M \longrightarrow S: $&$ M, S, \{K_{ms}\}_{K_c}, \{H(S, R, P)\}_{K_{ms}},$\\
                                 $                                                                                                                                                             
$&$ \{K_{ms}, H(S, R), T\}_{{K_m}^{-1}}$\\
(M3) $S \longrightarrow M: $&$ S, R, \{P\}_{K_{ms}}, \{K_{ms}, H(S, R), T\}_{{K_m}^{-1}}$\\
\end{tabular}\newline\newline
The key difference is the use of $\{K_{ms}, H(S, R), T\}_{{K_m}^{-1}}$ (which can only be generated by 
$M$) as the authorization token for verifying a response. $M$ checks whether $S$ is authorized by looking 
at $\{K_{ms}, H(S, R), T\}_{{K_m}^{-1}}$. Using this token, $S$ can just send message 3 alone, including 
a new text ($P'$), within the lifetime of $T$:\newline\newline 
\begin{tabular}{ll}
(M1) $S \longrightarrow M: $&$ S, R, \{P'\}_{K_{ms}}, \{K_{ms}, H(S, R), T\}_{{K_m}^{-1}}$\\
\end{tabular}\newline\newline
In these protocols, however, $S$ cannot find out where the challenge comes from. In an attempt to solve 
this problem, we assume there is an authenticated channel between $M$ to $S$, and $M$'s public key 
${K_m}$ is securely installed in a mobile device owned by $S$; perhaps during the process of 
manufacturing. 
We describe the following two protocols based on these assumptions: \newline\newline
\textbf{[Protocol 3]}\\
\begin{tabular}{ll}
(M1) $S \longrightarrow M: $&$ S, R, P$\\
(M2) $M \longrightarrow S: $&$ M, S, \{\{K_{ms}\}_{K_c}, N\}_{{K_m}^{-1}}$\\
(M3) $S \longrightarrow M: $&$ S, R, \{N + 1\}_{K_{ms}}$\\
\end{tabular}\newline\newline
In protocol 3, $M$ maintains the session information, \newline $(S, R, P, K_{ms}, N)$. When message 2 
arrives, $S$ verifies \newline the signature on $\{\{K_{ms}\}_{K_c}, N\}_{{K_m}^{-1}}$. $S$ does not 
\newline respond if the signature is unknown.\newline\newline
\textbf{[Protocol 4]}\\
\begin{tabular}{ll}
(M1) $S \longrightarrow M: $&$ S, R, P$\\
(M2) $M \longrightarrow S: $&$ M, S, \{\{K_{ms}\}_{K_c}, H(S, R), T\}_{{K_m}^{-1}},$\\
		 $										 $&$ \{P\}_{{K_m}^{-1}}$\\
(M3) $S \longrightarrow M: $&$ S, R, \{P\}_{K_{ms}},$\\
		 $										 $&$ \{\{K_{ms}\}_{K_C}, H(S, R), T\}_{{K_m}^{-1}}$\\
\end{tabular}\newline\newline
Protocol 4 uses $\{\{K_{ms}\}_{K_C}, H(S, R), T\}_{{K_m}^{-1}}$ as the authorized token. Our protocols 
are likely to be compatible with existing devices since the majority already have built-in encryption and 
hash functions.

\subsection{Observations}

\subsubsection{Upgrading Protocols}

A message is always sent to the message center of the contracted operator first. If the message is 
directed at someone contracted to a different operator, it is forwarded to another message center before 
reaching the receiver's handset~\cite{Enck05:SMS}. This means if one of the message centers decides not 
to use our framework, all uncertain texts delivered via that center would bypass the spam filter. It 
would be the weakest point (and the only route needed) for an attack. Hence, all existing message centers 
would have to support the new protocol. While this is a large change and a challenging one, 
operator-sponsored forums like OMTP (Open Mobile Terminal Platform), are working with key mobile 
operators to unify and recommend mobile terminal requirements~\cite{Roger07:OMTP}. With the increasing 
number of spam texts, it seems likely that the ability to filter machine-generated uncertain texts will 
persuade operators into upgrading their systems.

\subsubsection{Performance}

If there are too many messages subject to challenge-response, its overhead will dominate; for example, 
sending an image CAPTCHA is a huge overhead to authenticate a 100 character text message. Future work may 
look at adding a `bypass' to the hybrid: for example, if a message begins with a user-settable password 
(typically the recipient's name, but changeable), then it should be automatically treated as normal. As 
the uncertain region becomes smaller, we expect the performance of our framework to improve.

\subsubsection{Usability Issues}

Adapting CAPTCHA methods will have implications on the usability. A device might not have the capability 
to display an image CAPTCHA to a readable standard; also a mobile user might find it difficult to verify 
an audio CAPTCHA due to the background noise. These issues however, are likely to be resolved with 
technological advances. According to the Kelsey Group's second annual study on mobile use, more users 
than ever own internet-enabled smartphones: phones which provide advanced information accessing 
functions.

\section{Evaluation}
\label{section: Evaluation}

\subsection{Description of Datasets}
\begin{figure}[!h]
\centering
\begin{tabular}{c}
\includegraphics[scale=0.38]{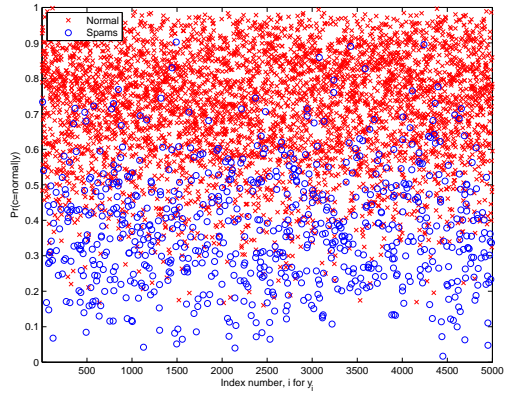}\cr
(a) N messages \cr
\includegraphics[scale=0.38]{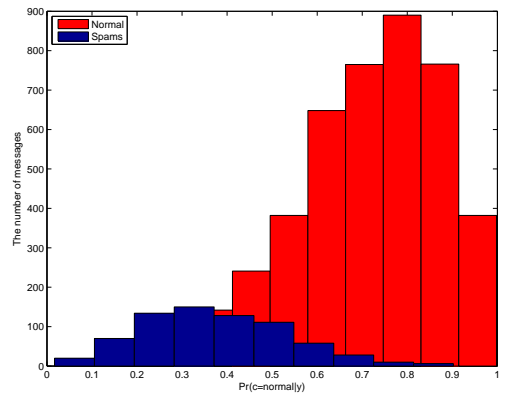}\cr
(b) Distribution
\end{tabular}
\caption{Displaying $\kappa=Pr(c=normal|y)$ for $N$ messages: spam (14.57\%) and normal (85.32\%)}
\label{fig: Displaying kappa for N messages}
\end{figure}

In order to measure the performance of our framework, we have generated synthetic datasets. Suppose that 
there are $N$ number of sent messages (we set $N=5000$). We will use $p$ and $q$ to show the normal to 
spam proportion where $p+q=1$, and $p$ and $q$ are non-negative numbers (in reality, there would be many 
operators with different proportions). Let $\kappa$ be a random variable generated from an existing 
filtering method: $\kappa=Pr(c=\emph{normal}|y)$. For an artificial dataset, we build a mixture model 
given by
\begin{eqnarray}
p(\kappa|\lambda)&=&p(\kappa|c=normal, \lambda)p(c=normal|\lambda)\nonumber\\
				 &&+p(\kappa|c=spam, \lambda)p(c=spam|\lambda)
\end{eqnarray}
where $\lambda$ denotes a set of hyper-parameters which control parameters. We assume $c_{i}$ is 
generated from binomial distribution with hyper-parameters $p$ and $q$. Thus, we have:
\[
c\sim p(c|\lambda)=Pr(c|1, p)=p^{c}(1-p)^{1-c}=p^{c}q^{1-c}
\]
After classifying the $i$th sample message, we generate the expected probability (this is the filtering 
output):
\[
\kappa \sim p(\kappa|c, \lambda)=
\left\{
\begin{matrix}
p(\kappa|c=normal, \lambda)&=&\mathcal{B}(\kappa; \alpha_{1}, \beta_{1})\cr
p(\kappa|c=spam, \lambda)&=&\mathcal{B}(\kappa; \alpha_{0}, \beta_{0})\cr
\end{matrix}
\right.
\]
Here, $\mathcal{B}$ represents beta distribution and its hyper-parameters are set as follows: 
$\alpha_{0}=3$, $\beta_{0}=5$, $\alpha_{1}=5$, $\beta_{1}=2$. Both thresholds ($h_{1}$ and $h_{2}$) vary 
between $0$ and $1$ by $1/30$.

We have built an artificial dataset based on a Spanish database \cite{Hidalgo06:Spam} which shows the 
proportion of spam as 14.57\% and normal as 85.32\%; that is, $q=0.1457$ and $p=0.8532$. The generated 
data have been plotted in Fig. \ref{fig: Displaying kappa for N messages}-(a). A red cross represents 
normal message and a blue circle represents spam. This colouring scheme is also used in Fig. \ref{fig: 
Displaying kappa for N messages}-(b). The graphs show that there are a lot of overlapping labels between 
$0.2$ and $0.8$. This overlapping section is considered as the uncertain region. Note that the 
challenge-response is not perfect and some of the spam might bypass the filter with correct responses, 
and normal messages might be filtered with incorrect ones. To model this imperfection, we use $e_{1}$ and 
$e_{2}$ to represent the ratios of False Positives (FP) and False Negatives (FN) in the uncertain region.

\subsection{Traffic Usage Comparison}
We have simulated and analyzed the traffic usage using the variable thresholds. Our framework considers 
three major stakeholders (see Fig. \ref{fig: frameworks in mobile system}): a message sender (A), a 
message center (B), and a receiver (C).

\comment{
\begin{figure}[!h]
\centering
\includegraphics[scale=0.25]{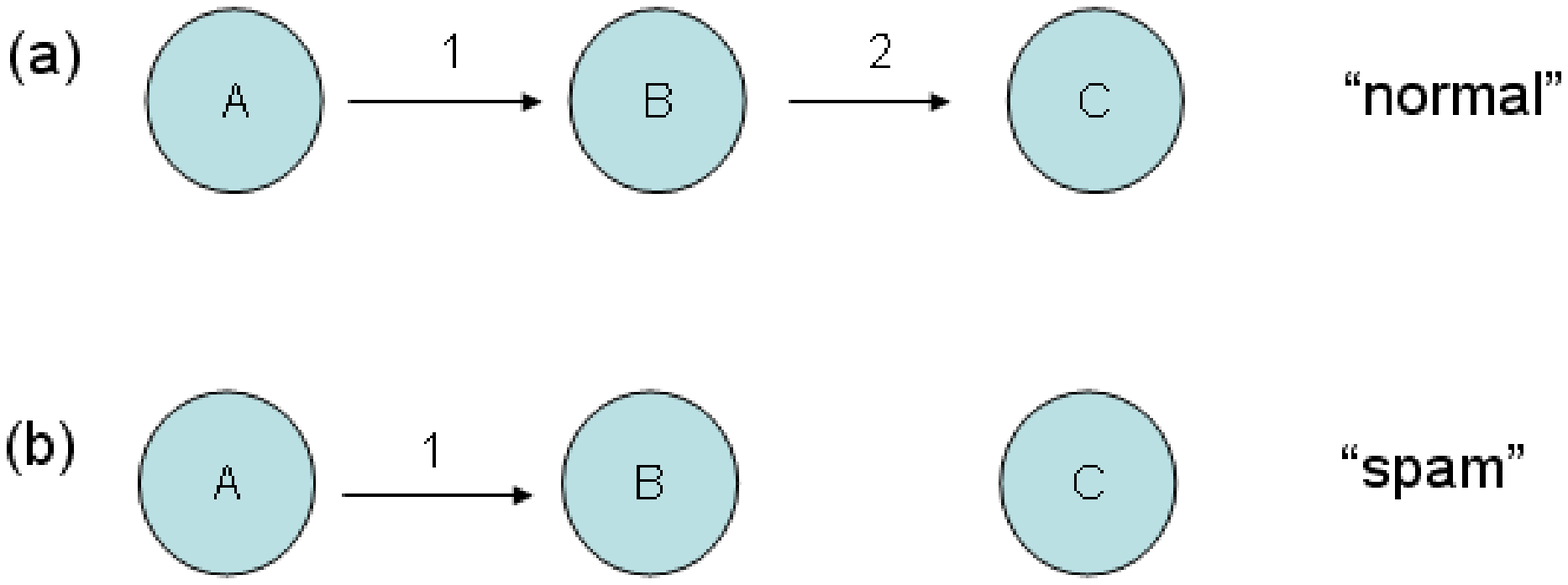}
\caption{Two possible pathways for the filtering method: (a) a normal message reaches C via B (b) spam is 
deleted at B}
\label{fig: two possible cases for filtering only approach}
\end{figure}
}

First, we calculate the traffic used by an existing filtering method. Only the messages with filtering 
probabilities higher than the threshold $h$ reach C via B; other messages are deleted at B (only A to B). 
Suppose that ${\bf y}_{\bar{c}=type}^{h}$ for $type\in\{normal, spam\}$ denotes all messages filtered as 
$type$ in terms of $h$, the total amount of traffic used is the sum of $|{\bf 
y}_{\bar{c}=normal}^{h}|\times 2$, and $|{\bf y}_{\bar{c}=spam}^{h}|\times 1$ where $|\cdot|$ represents 
the cardinality of a set: $N_{Filtering Only}=|{\bf y}_{\bar{c}=normal}^{h}|\times 2 + |{\bf
y}_{\bar{c}=spam}^{h}|\times 1$. This is because a normal message has two paths (A$\rightarrow$B and 
B$\rightarrow$C) but spam only has one path (A$\rightarrow$B).
\comment{
 as shown in Fig. \ref{fig: two possible cases for filtering only approach}.
 } 
\begin{figure}[!h]
\centering
\includegraphics[scale=0.35]{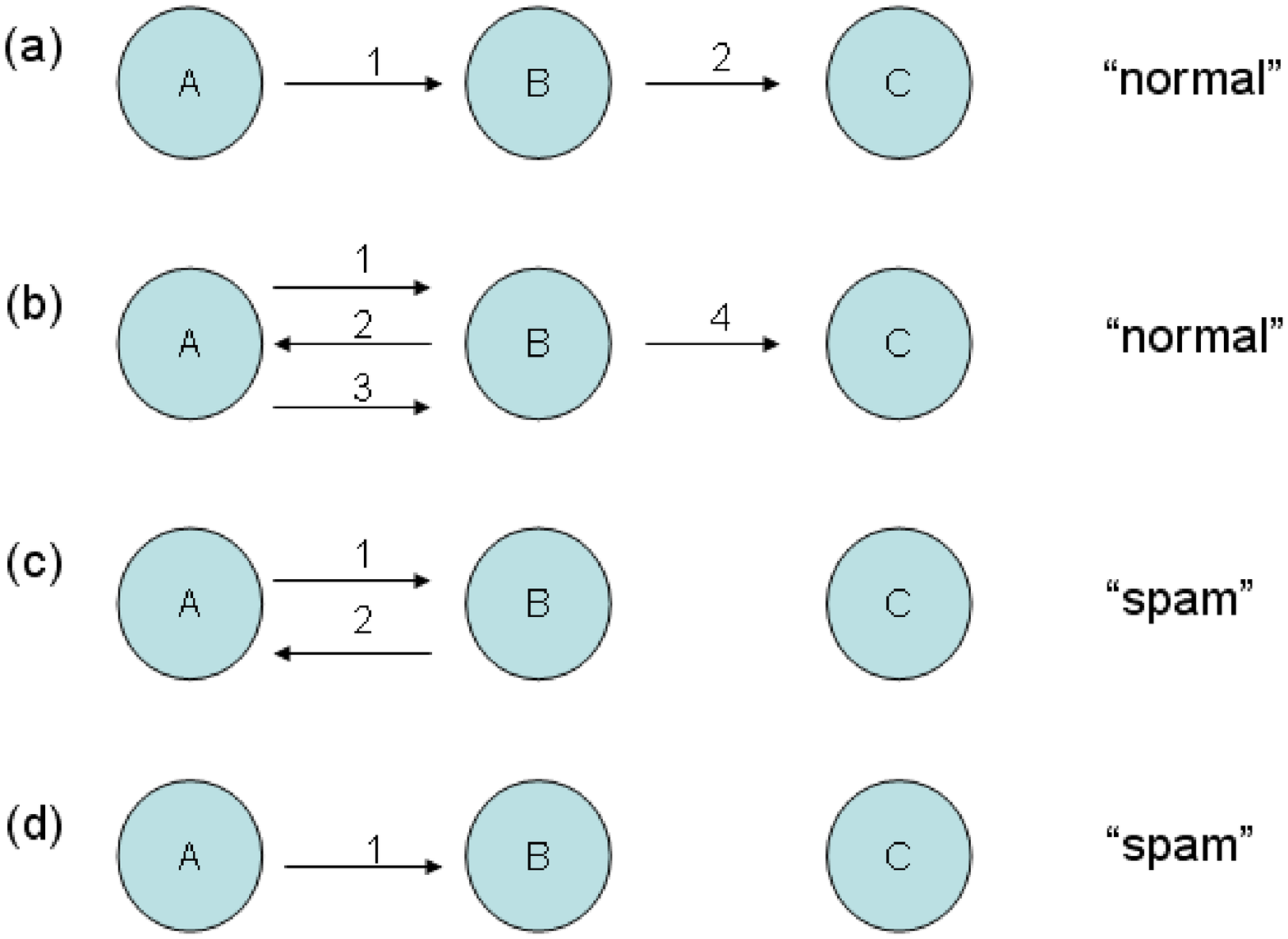}
\caption{Four possible pathways for the hybrid method}
\label{fig: 4 possible cases for the hybrid approach}
\end{figure}
In contrast, our hybrid model divides the measurable space into three different areas using two 
thresholds: $h_{1}$ and $h_{2}$. As a result, we have two more parameters to estimate: the traffic used 
by normal messages ($N_{un}$) and spam ($N_{us}$) in the uncertain region. Let ${\bf y}_{\tilde{c}=type}$ 
for $type\in\{normal, spam\}$ be a set of messages that have label $type$ as a ground truth. 

As Fig. \ref{fig: 4 possible cases for the hybrid approach} shows, there are four possible pathways. 
Firstly, a message classified as normal using the higher threshold is sent directly to C via B; the 
number of paths taken is two ($A\rightarrow B\rightarrow C$). Secondly, a message is in between the 
higher and the lower thresholds; a correct response is submitted by the sender and the message is 
classified as normal; the number of paths taken is four ($A\rightarrow B \rightarrow A \rightarrow B 
\rightarrow C$). Thirdly, a message is again in between two thresholds; this time no response is returned 
and the message is classified as spam; the number of paths taken is two ($A\rightarrow B \rightarrow A$). 
Lastly, a message classified as spam using the lower threshold is deleted at B; the number of paths taken 
is one ($A\rightarrow B$).

The traffic usage is calculated using:
\begin{eqnarray}
N_{n} &=& |{\bf y}_{\bar{c}=\emph{normal}}^{h_{2}}|\times 2\nonumber\\
N_{un} &=& |{\bf y}_{\bar{c}=\emph{normal}}^{h_{1}}\cap {\bf y}_{\bar{c}=\emph{spam}}^{h_{2}}\cap {\bf 
y}_{\tilde{c}=\emph{normal}}|\times (1-e_{1}) \times 4\nonumber\\
		&&+|{\bf y}_{\bar{c}=\emph{normal}}^{h_{1}}\cap {\bf y}_{\bar{c}=\emph{spam}}^{h_{2}}\cap {\bf 
y}_{\tilde{c}=\emph{spam}}|\times e_{2} \times 4\nonumber\\
N_{us} &=& |{\bf y}_{\bar{c}=\emph{normal}}^{h_{1}}\cap {\bf y}_{\bar{c}=\emph{spam}}^{h_{2}}\cap {\bf 
y}_{\tilde{c}=\emph{spam}}|\times (1-e_{2}) \times 2\nonumber\\
		&&+|{\bf y}_{\bar{c}=\emph{normal}}^{h_{1}}\cap {\bf y}_{\bar{c}=\emph{spam}}^{h_{2}}\cap {\bf 
y}_{\tilde{c}=\emph{normal}}|\times e_{1} \times 2\nonumber\\
N_{s} &=& |{\bf y}_{\bar{c}=\emph{spam}}^{h_{1}}|\times 1\nonumber\\
N_{hybrid} &=& N_{n}+N_{un}+N_{us}+N_{s}
\end{eqnarray}
where $e_{1}$ and $e_{2}$ are the probability which normal people cannot respond. Here, $e_{1}$ and 
$e_{2}$ are fixed to values $0.02$ and $0.01$ respectively. 

\comment{
Fig. \ref{fig: 3D view in terms of varying thresholds} shows the traffic usage with varying thresholds, 
$h_{1}$ and $h_{2}$. We assume that $h_{1}$ is smaller than $h_{2}$, and only the right-half of the graph 
is meaningful. The green plane represents the traffic usage for using the filtering method and the blue 
represents the usage for our hybrid framework.
\begin{figure}[!h]
\centering
\includegraphics[scale=0.4]{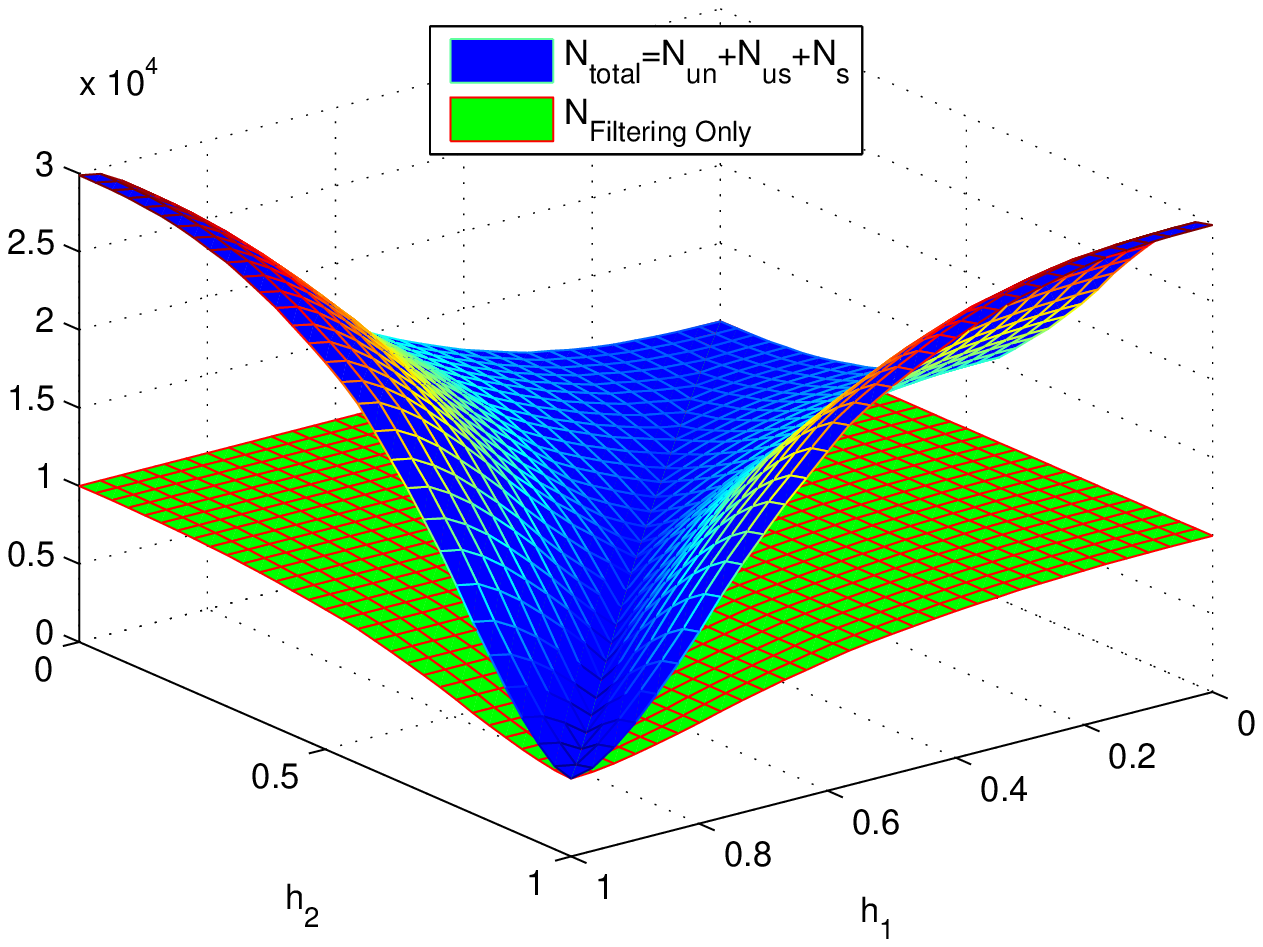}
\caption{3D view in terms of two varying thresholds}
\label{fig: 3D view in terms of varying thresholds}
\end{figure}
}
Figures \ref{fig: Slides of an axis}-(a) and \ref{fig: Slides of an axis}-(b) show the inner sections of 
the graph. The higher threshold is fixed to $0.73333$, only the lower threshold increases from $0$ until 
it reaches this value. The traffic is used less when the lower threshold increases.
\begin{figure}[!h]
\centering
\begin{tabular}{c}
\includegraphics[scale=0.47]{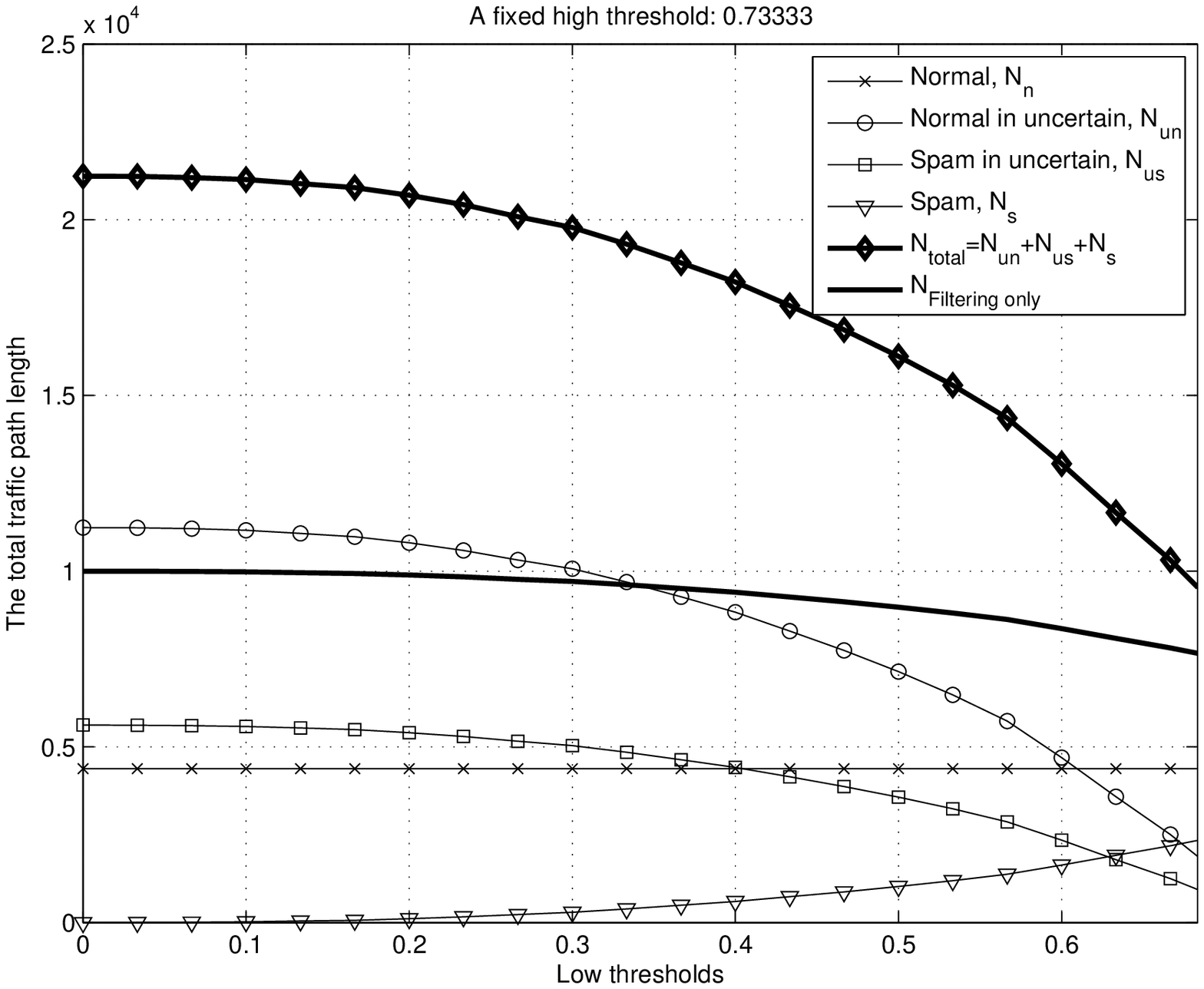}\cr
(a) A high threshold\cr
\includegraphics[scale=0.47]{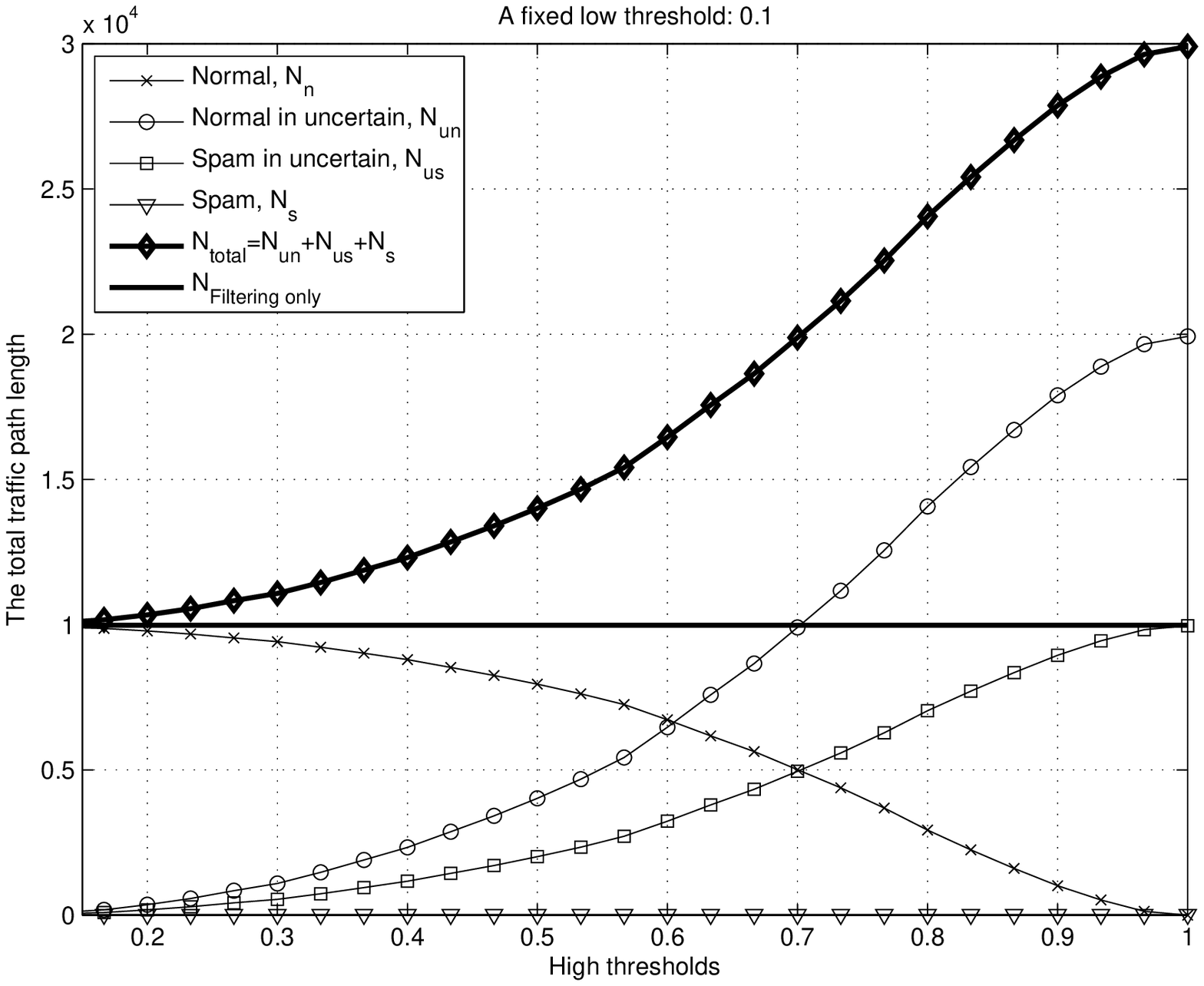}\cr
(b) A low threshold
\end{tabular}
\caption{Slides of an axis (with fixed threshold)}
\label{fig: Slides of an axis}
\end{figure}
We have also monitored the traffic usage with the lower threshold fixed to $=0.1$, and with the higher 
threshold increased from this value to $1$ (see Fig.\ref{fig: Slides of an axis}-a). The traffic usage 
does not change with the filtering approach because the lower threshold is the same as $h$. As the number 
of messages in the uncertain region increases so does the traffic usage.  
\subsection{Variant proportion of spam}

We have fixed the proportion of spam to $14.57\%$ and normal messages to $85.32\%$. In this section we 
show how the performance is affected when these proportions change. 

Table \ref{Table: Traffic Amounts in terms of thresholds} describes a small number of samples from the 
nine different proportions. Each record has six different columns: proportion of spam ($\%$), lower 
threshold ($h_{1}$), higher threshold ($h_{2}$), traffic usage (TA) of $N_{hybrid}$, ratio 
$\left(=\frac{N_{hybrid}}{N_{filtering only}}\right)$, and accuracy $\left(ACC=\frac{TP+TN}{P+N}\right)$. 
It uses three different measures for the performance. If the traffic usage is less, we say the system is 
lighter and is more economical. The ratio is only close to $1$ if the traffic used in the hybrid method 
is close to the amount used in the other. The accuracy measures the correctness of message 
classification. We can select practical threshold values for each spam proportion to compare the 
performance. For instance, threshold values $h_{1}=0.1$ and $h_{2}=0.2$ can be selected in $10\%$ spam 
proportion to show a reasonable performance of the hybrid method.
However, if the system is concerned with achieving high accuracy and not with reducing the traffic usage, 
$h_{1}=0.1$ and $h_{2}=0.9$ values can be used. In a spam-dominant environment (for spam proportion of 
$50\%$), reasonable threshold values would be $h_{1}=0.4$ and $h_{2}=0.6$. Returning back to the figures 
for a spam proportion of $10\%$, $h_{1}=0.1$ and $h_{2}=0.9$ will be selected when accuracy is the most 
significant factor. 
\begin{table}[!t]
\caption{Traffic amounts and accuracy of hybrid methods in terms of thresholds}
\label{Table: Traffic Amounts in terms of thresholds}
\centering
\begin{tabular}{||c| c|c| c| c| c||}
\hline\hline
Proportion 	& $h_{1}$ 	& $h_{2}$ 	& TA	&	Ratio 	& ACC\\
of spam 	& & & &	& \\\hline\hline
			&	0.1		& 	0.2	& 	10218.8	&	1.0239	& 0.91847 \\
			&	0.1		& 	0.9	& 	27698.96&	2.7754	& 0.98312 \\
			&	0.4		& 	0.6	& 	13563.78&	1.4218	& 0.95266 \\[-3ex]
\raisebox{3ex}{10\%}			&	0.8		& 	0.9	& 	10493.06&	1.6081	& 0.39682\\[0.5ex]\hline
			&	0.1		& 	0.2	& 	10450.56&	1.0487	& 0.83314\\
			&	0.1		& 	0.9	& 	27859.08&	2.7957	& 0.98391\\
			&	0.4		& 	0.6	& 	13561.34&	1.4659	& 0.94151\\[-3ex]
\raisebox{3ex}{20\%}			&	0.8		& 	0.9	& 	10050.6	&	1.5731	& 0.46933\\[0.5ex]\hline
			&	0.1		& 	0.2	& 	10662.46&	1.0707	& 0.74703\\
			&	0.1		& 	0.9	& 	28114.76&	2.8233	& 0.98462\\
			&	0.4		& 	0.6	& 	13682.54&	1.5179	& 0.94212\\[-3ex]
\raisebox{3ex}{30\%}			&	0.8		& 	0.9	& 	9404.9	&	1.5167	& 0.53119\\[0.5ex]\hline
			&	0.1		& 	0.2	& 	10915.14&	1.0972	& 0.6635\\
			&	0.1		& 	0.9	& 	28188.34&	2.8336	& 0.9853\\
			&	0.4		& 	0.6	& 	13717.14&	1.5641	& 0.93292\\[-3ex]
\raisebox{3ex}{40\%}			&	0.8		& 	0.9	& 	8721.52	&	1.4442	& 0.59632\\[0.5ex]\hline
			&	0.1		& 	0.2	& 	11139.94&	1.1214	& 0.58259\\
			&	0.1		& 	0.9	& 	28341.5	&	2.853	& 0.98627\\
			&	0.4		& 	0.6	& 	13523.66&	1.5946	& 0.92748\\[-3ex]
\raisebox{3ex}{50\%}			&	0.8		& 	0.9	& 	8182.46	&	1.3902	& 
0.66251\\[0.5ex]\hline\hline
\end{tabular}
\end{table}
\section{Conclusion and Future Work}
\label{section: conclusion}
We proposed a hybrid spam filtering framework for mobile communication using a combination of 
\emph{content-based filtering} and \emph{challenge-response}. A message that falls into the uncertain 
region (after filtering), is further classified by sending a challenge (e.g. an image CAPTCHA) to the 
sender: a legitimate sender is likely to answer it correctly, whereas an automated spam program is not. 
Challenge-response protocols have been designed with the necessary cryptographic features.  We have also 
shown the trade-offs between the \emph{accuracy} and the \emph{traffic usage} in using our framework. The 
simulation results suggest that, for a different level of spam proportion, the practical thresholds 
should be carefully selected according to the required level of the two measures.

In this paper, a synthetic dataset, as oppose to a real dataset has been used due to the following three 
reasons: firstly, we wanted to develop a generalized framework that is flexible and applicable to a wide 
range of applications (e.g VoIP spam filters\cite{Croft05:VOIPspam}); secondly, it was not easy to find a 
real dataset since the challenge-response protocol is not a commonly used filtering method; lastly, this 
protocol involves a great level of human interaction and developing such a prototype (in order to 
generate our own dataset) was outside the scope. Our next step will be to contact mobile operators and 
forums like OMTP to collect real data, and evaluate our framework against other datasets.

Having the network operators charge for sending of text messages has been one of the big inhibitors to 
the growth of spam: even a cent per message might hugely alter the economics of a spammer. Assuming that 
a reasonable filtering method is in place, another hybrid potential is to force spammers to opt into a 
charging scheme where the cost of responding to a challenge is larger than sending an initial spam. For 
example, imagine it costs two cents to send a spam, then it would cost extra five cents to answer an 
image CAPTCHA.    

\comment{
We have described a simplified model involving only three stakeholders, a sender, a message center and a 
receiver. In reality, there will be more involved; for instance, there might be a base station, or 
several message centers owned by different operators. We will expand our model to consider other 
stakeholders and derive more accurate simulation results based on it. This model will also look at 
different types of network available, such as, GSM, UMTS or CDPD.   

\section*{Acknowledgements}
The authors would like to thank Ross Anderson and Andrew Martin for their careful attention and 
insightful comments. 
}

\bibliographystyle{unsrt}
\bibliography{HybridSpamTech}
\end{document}